\documentclass[12pt,preprint2]{aastex}

  \usepackage{subfigure} 
\usepackage{graphicx}
\usepackage{epsf}

\newcommand{\bef}{\begin{figure}}
\newcommand{\eef}{\end{figure}}

\def\eps@scaling{0.95}

\def\showone#1{
  \centering
  \leavevmode
  \epsfxsize=\eps@scaling\linewidth
  \epsfbox{#1.eps}
}

\def\showtwover#1#2{
  \centering
  \leavevmode
  \epsfxsize=\eps@scaling\linewidth
  \epsfbox{#1.eps} \hfil
  \epsfxsize=\eps@scaling\linewidth
  \epsfbox{#2.eps}
}

\def\showthreeover#1#2#3{
  \centering
  \leavevmode
  \epsfxsize=\eps@scaling\linewidth
  \epsfbox{#1.eps} \hfil
  \epsfxsize=\eps@scaling\linewidth
  \epsfbox{#2.eps} \hfil
  \epsfxsize=\eps@scaling\linewidth
  \epsfbox{#3.eps}
}

\def\showfourover#1#2#3#4{
  \centering
  \leavevmode
  \epsfxsize=\eps@scaling\linewidth
  \epsfbox{#1.eps} \hfil
  \epsfxsize=\eps@scaling\linewidth
  \epsfbox{#2.eps} \hfil
  \epsfxsize=\eps@scaling\linewidth
  \epsfbox{#3.eps} \hfil
  \epsfxsize=\eps@scaling\linewidth
  \epsfbox{#4.eps}
}

\def\epstwo@scaling{0.46}

\def\showtwo#1#2{
  \centering
  \leavevmode
  \epsfxsize=\epstwo@scaling\linewidth
  \epsfbox{#1.eps} 
  \epsfxsize=\epstwo@scaling\linewidth
  \epsfbox{#2.eps}
}

\def\epsthree@scaling{0.28}
\def\showthree#1#2#3{
  \centering
  \leavevmode
  \epsfysize=\epsthree@scaling\textwidth %% \linewidth
  \epsfbox{#1.eps} 
  \epsfysize=\epsthree@scaling\textwidth %% \linewidth
  \epsfbox{#2.eps}
  \epsfysize=\epsthree@scaling\textwidth %% \linewidth
  \epsfbox{#3.eps}
}

\def\epstwo@scaling{0.44}

\def\showfour#1#2#3#4{
  \centering
  \leavevmode
  \epsfxsize=\epstwo@scaling\linewidth
  \epsfbox{#1.eps} \hfil
  \epsfxsize=\epstwo@scaling\linewidth
  \epsfbox{#2.eps} \hfil
  \epsfxsize=\epstwo@scaling\linewidth
  \epsfbox{#3.eps} \hfil
  \epsfxsize=\epstwo@scaling\linewidth
  \epsfbox{#4.eps}
}

\def\showsix#1#2#3#4#5#6{
  \centering
  \leavevmode
  \epsfxsize=\epstwo@scaling\linewidth
  \epsfbox{#1.eps} \hfil
  \epsfxsize=\epstwo@scaling\linewidth
  \epsfbox{#2.eps} \hfil
  \epsfxsize=\epstwo@scaling\linewidth
  \epsfbox{#3.eps} \hfil
  \epsfxsize=\epstwo@scaling\linewidth
  \epsfbox{#4.eps} \hfil
  \epsfxsize=\epstwo@scaling\linewidth
  \epsfbox{#5.eps} \hfil
  \epsfxsize=\epstwo@scaling\linewidth
  \epsfbox{#6.eps}
}

\newcommand{\befone}{
  \begin{figure*}
  \centering
  \begin{minipage}{\textwidth}
  }
\newcommand{\eefone}{\end{minipage}\end{figure*}}

\newcommand{\sm}[1]{\mbox{{\scriptsize #1}}}

\newcommand{\HI}{$\mathrm{H}$\ }
\newcommand{\HII}{$\mathrm{H}^+$\ }

\newcommand{\HeI}{$\mathrm{He}$\ }

\newcommand{\fHI}{\mathrm{H}}
\newcommand{\fHII}{\mathrm{H}^+}

\newcommand{\fHeI}{\mathrm{He}}

\newcommand{\fe}{\mathrm{e}}
\newcommand{\fpp}{\mathrm{p}}

\shorttitle{Primordial magnetic fields and 21 cm emission}
\shortauthors{Schleicher et al.}

\begin{document}


%\title{Magnetic fields in the early universe.\\ II. Implications for the $21$ cm measurements}
%\title{Imprints of primordial magnetic fields in the 21 cm line.}
%\title{Observational consequences of primordial magnetic fields with respect to $21$ cm measurements.}
\title{Influence of primordial magnetic fields on 21 cm emission}


\author{Dominik R. G. Schleicher, Robi Banerjee, Ralf S. Klessen}
\affil{Zentrum f\"ur Astronomie der Universit\"at Heidelberg, \\Institut f\"ur Theoretische Astrophysik,\\ Albert-Ueberle-Str. 2,\\ D-69120 Heidelberg,\\ Germany}
\email{dschleic@ita.uni-heidelberg.de, banerjee@ita.uni-heidelberg.de, rklessen@ita.uni-heidelberg.de}

%\author{}
%\affil{Institute of Theoretical Astrophysics / ZAH, Albert-Ueberle-Str. 2, D-69120 Heidelberg, Germany}
%\email{}

%\and

%\author{}
%\affil{Institute of Theoretical Astrophysics / ZAH, Albert-Ueberle-Str. 2, D-69120 Heidelberg, Germany}
%\email{}


%\altaffiltext{1}{Visiting Astronomer, Cerro Tololo Inter-American Observatory.
%CTIO is operated by AURA, Inc.\ under contract to the National Science
%\altaffiltext{2}{Society of Fellows, Harvard University.}
%\altaffiltext{3}{present address: Center for Astrophysics,
%    60 Garden Street, Cambridge, MA 02138}
%\altaffiltext{4}{Visiting Programmer, Space Telescope Science Institute}
%\altaffiltext{5}{Patron, Alonso's Bar and Grill}


\begin{abstract}
Magnetic fields in the early universe can significantly alter the thermal evolution and the ionization history during the dark ages. This is reflected in the $21$ cm line of atomic hydrogen, which is coupled to the gas temperature through collisions at high redshifts, and through the Wouthuysen-Field effect at low redshifts. We present a semi-analytic model for star formation and the build-up of a Lyman $\alpha$ background in the presence of magnetic fields, and calculate the evolution of the mean $21$ cm brightness temperature and its frequency gradient as a function of redshift. We further discuss the evolution of linear fluctuations in temperature and ionization in the presence of magnetic fields and calculate the effect on the $21$ cm power spectrum. At high redshifts, the signal is increased compared to the non-magnetic case due to the additional heat input into the IGM from ambipolar diffusion and the decay of MHD turbulence.  At lower redshifts, the formation of luminous objects and the build-up of a Lyman $\alpha$ background can be delayed by a redshift interval of $10$ due to the strong increase of the filtering mass scale in the presence of magnetic fields. This tends to decrease the $21$ cm signal compared to the zero-field case. In summary, we find that $21$ cm observations may become a promising tool to constrain primordial magnetic fields.
\end{abstract}





\section{Introduction}
{Observations of the $21$ cm fine structure line of atomic hydrogen have the potential to become an important means of studying} the universe at early times, during and even before 
the epoch of reionization. This possibility was suggested originally by \citet{Purcell}, and significant
process in instrumentation and the development of radio telescopes has brought us close to the first observations
from radio telescopes like LOFAR\footnote{http://www.lofar.org}. While one of the main purposes is to increase our 
understanding of cosmological reionization \citep{Bruyn}, a number of further exciting applications have been suggested
in the mean time. \citet{Loeb} demonstrated how $21$ cm measurements can probe the thermal evolution of the IGM at a
much earlier time, at redshifts of $z\sim200$. \citet{Barkana} suggested a method that separates physical and astrophysical 
effects and thus allows to probe the physics of the early universe. \citet{FurlanettoDM} showed how the effect of
dark matter annihilation and decay would be reflected in the $21$ cm line, and effects of primordial magnetic fields have been considered
by \citet{Tashiro}.\\ \\

Indeed, primordial magnetic fields can affect the early universe in various ways. The thermal evolution is significantly altered by 
ambipolar diffusion heating and decaying MHD turbulence \citep{Sethi05, Sethi08, Schleicher}. \citet{Kim} calculated the effect of the Lorentz force on structure formation and showed that additional power is present on small scales in the presence of primordial magnetic fields. It was thus suggested that reionization occurs earlier in the presence of primordial magnetic fields \citep{Sethi05, TashiroReion}. However, as pointed out by \citet{Gnedin}, the characteristic mass scale of star forming halos, the so-called filtering mass, increases signficantly when the temperature is increased. For comoving field strengths of $\sim1$ nG, we found that the filtering mass scale is shifted to scales where the power spectrum is essentially independent of the magnetic field \citep{Schleicher}. Reionization is thus delayed in the presence of primordial magnetic fields. We further found upper limits of the order $1$ nG, based on the Thomson scattering optical depth measured by WMAP 5 \citep{Komatsu, WMAPAngular} and the requirement that reionization ends at $z\sim6$ \citep{Becker}. 

%While works of \citet{Abel}
%and \citet{Bromm} indicate the formation of very massive stars,
%\citet{Clark} showed that the collapsing gas in the first star forming
%halos may further fragment and form a stellar cluster due to a stage of
%efficient colling~\citep{Omukai05}. \citet{Omukai} suggested that the same is true for atomic-cooling halos. \citet{Silk} propose further that the presence of primordial magnetic fields may lead to a star formation scenario closer to present day star formation.

The presence of primordial magnetic fields can have interesting implications on first star formation as well. In the absence of magnetic fields, \citet{Abel} and \citet{Bromm} suggested that the first stars should be very massive, perhaps with $\sim100$ solar masses.  \citet{Clark} and \citet{Omukai} argued that in more massive and perhaps metal-enriched galaxies, fragmentation should be more effective and lead to the formation of rather low-mass stars, due to a stage of efficient cooling \citep{Omukai}. {Magnetic fields may change this picture and reduce the stellar mass by triggering jets and outflows \citep{Silk}. However, simulations by \citep{Machida} show that the change in mass is of the order $10\%$.} \\ \\

$21$ cm measurements can try to adress {primordial magnetic fields} in two ways: During the dark ages of the universe, at redshifts $z\sim200$ well before the formation of the first stars, the spin temperature of hydrogen is coupled to the gas temperature via collisional de-excitation by hydrogen atoms \citep{Allison, Zygelman} and free electrons \citep{Smith, FF07}, constituting a probe at very early times. While collisional de-excitation becomes inefficient due to the expansion of the universe, the first stars will build up a Lyman $\alpha$ background that will cause a deviation of the spin temperature from the radiation temperature by the Wouthuysen-Field effect \citep{Wouthuysen, Field}. As primordial magnetic fields may shift the onset of reionization, the onset of this coupling constitutes an important probe on the presence of such fields. We adress these possibilities in the following way: In \S\ref{IGM}, we review our treatment of the IGM in the presence of primordial magnetic fields. The evolution of the $21$ cm background and the role of Lyman $\alpha$ photons is discussed in \S\ref{21cm}. The evolution of linear perturbations in temperature and ionization is calculated in \ref{fluctuations}. The results for the power spectrum are given in \S\ref{power}, and the results are further discussed in \S\ref{discussion}.

\section{The evolution of the IGM}\label{IGM}
As indicated in the introduction, primordial magnetic fields can have
a strong impact on the evolution of the IGM before reionization. Once
the first luminous objects form, their feedback must also be taken
into account. In this section, we review the basic ingredients of our
treatment of the IGM between recombination and reionization. We refer
the interested reader to \citet{Schleicher} for more details.

\subsection{The RECFAST code}

% \begin{figure}
%   \centering
%    \subfigure[]{\includegraphics[scale=0.4]{f1.eps}\label{temperature}}\qquad
%    \subfigure[]{\includegraphics[scale=0.4]{f2.eps}\label{ionization}}\qquad
%  
%   \caption{Evolution of (a) temperature and (b) ionized fraction in the medium that was not yet affected by UV feedback. The details of the models are given in Table \ref{tab:models}. } 
% \end{figure}

\bef
\showtwover{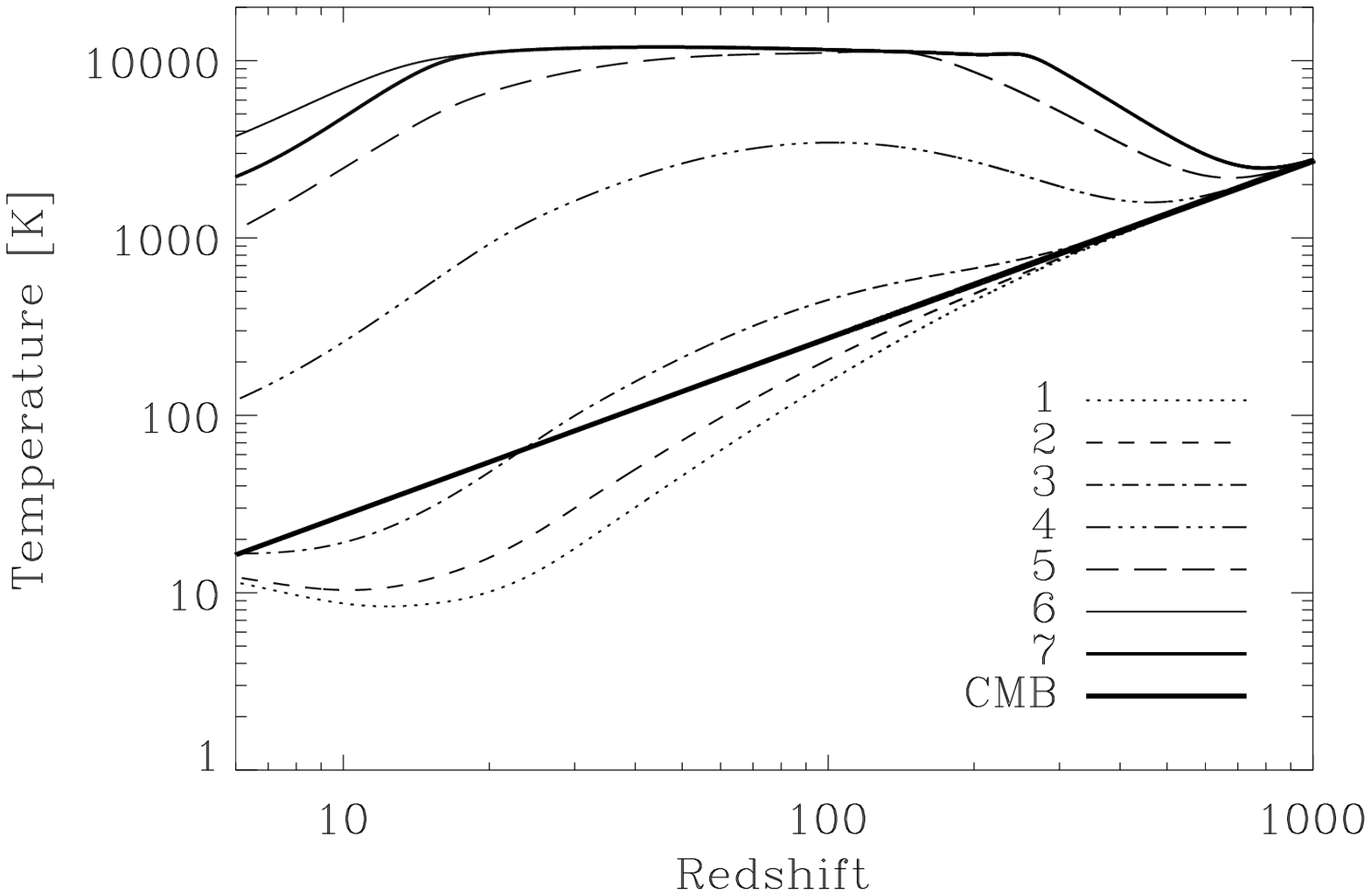}
	{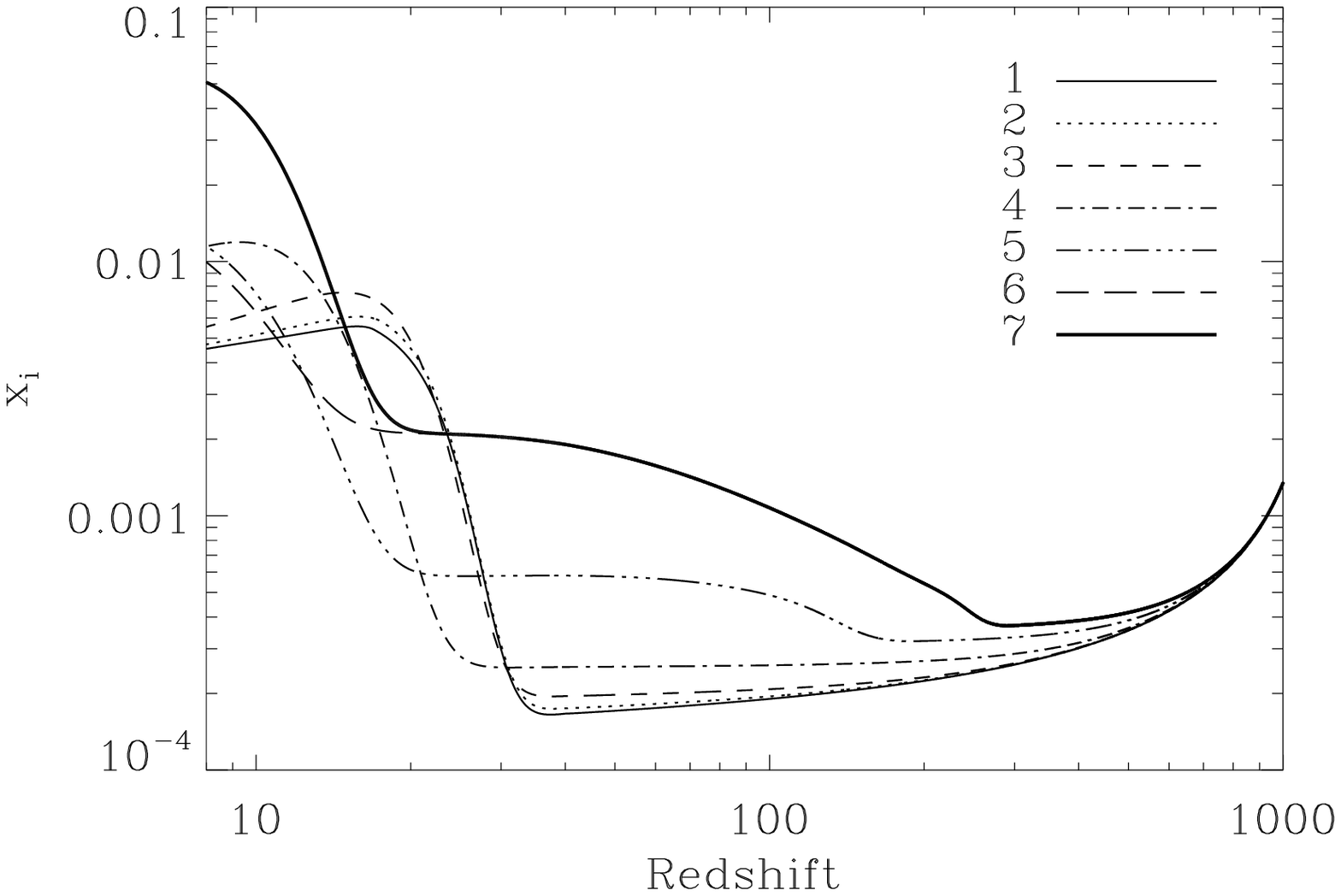}
\caption{Evolution of gas temperature (upper panel) and ionized fraction (lower panel) in the medium that was not yet affected by UV feedback. The details of the models are given in Table \ref{tab:models}.}
\label{tempion}
\eef

\begin{figure}
  \centering
  \includegraphics[scale=0.4]{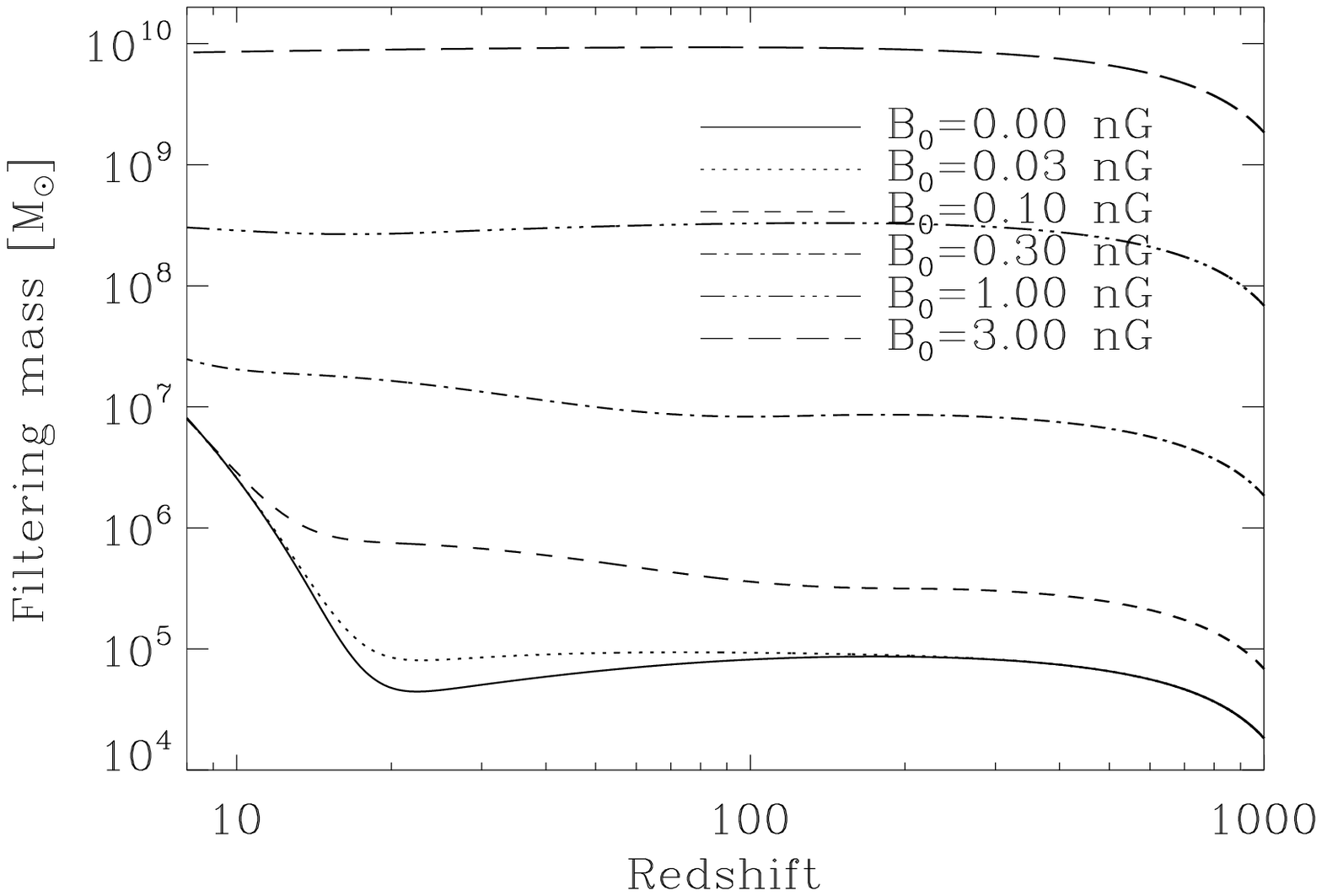}\label{filt}
 
  \caption{Evolution of the filtering mass scale for different comoving magnetic field strengths.} 
\end{figure}

Our calculation is based on a modified version of the RECFAST code\footnote{http://www.astro.ubc.ca/people/scott/recfast.html} \citep{Seager,SeagerFast} that calculates recombination and the freeze-out of electrons. The calculation of helium recombination was recently updated by \citet{Wong}. We have extended this code by including a model for reionization in the context of primordial magnetic fields \citet{Schleicher}. The equation for the temperature evolution is given as
\begin{eqnarray}
\frac{dT}{dz}&=&\frac{8\sigma_T a_R T_r^4}{3H(z)(1+z)m_e c}\frac{x_e}{1+f_{\fHeI}+x_e}(T-T_r)\nonumber\\
&+&\frac{2T}{1+z}-\frac{2(L_{\sm{heat}}-L_{\sm{cool}})}{3nk_B H(z)(1+z)},\label{temp}
\end{eqnarray}
where $L_{\sm{heat}}$ is the new heating term (see \S \ref{magnetic}, \S \ref{stellar} and \S \ref{Lymanbg}), $L_{\sm{cool}}$ the cooling by Lyman $\alpha$ emission, bremsstrahlung and recombinations, $\sigma_T$ is the Thomson scattering cross section, $a_R$ the Stefan-Boltzmann radiation constant, $m_e$ the electron mass, $c$ the speed of light, $k_B$ Boltzmann's constant, $n$ the total number density, $x_e=n_e/n_{\fHI}$ the electron fraction per hydrogen atom, $H(z)$ is the Hubble factor and $f_{\fHeI}$ is the number ratio of \HeI and \HI nuclei, which can be obtained as $f_{\fHeI}=Y_p/4(1-Y_p)$ from the mass fraction $Y_p$ of \HeI with respect to the total baryonic mass. {Eq. (\ref{temp}) describes the change of temperature with redshift due to Compton scattering of CMB photons, expansion of the universe  and additional heating and cooling terms that are described in the following subsections in more detail.} The evolution of the ionized fraction of hydrogen, $x_\fpp$, is given as
\begin{eqnarray}
\frac{dx_\fpp}{dz}&=&\frac{[x_\fe x_\fpp n_\fHI \alpha_\fHI
    -\beta_\fHI (1-x_\fpp)e^{-h_p\nu_{\fHI, 2s}/k_B T}] }{H(z)(1+z)[1+K_\fHI(\Lambda_\fHI+\beta_\fHI)n_\fHI(1-x_\fpp)]}\nonumber\\
&\times&[1+K_\fHI\Lambda_\fHI n_\fHI(1-x_\fpp)]\nonumber\\
&-&\frac{k_{\sm{ion}}n_H x_\fpp}{H(z)(1+z)}-f_{\sm{ion}},\label{ion}
\end{eqnarray}
{where the ionization fraction is determined by radiative recombination, photo-ionization of excited hydrogen atoms by CMB photons in an expanding universe and collisional ionization, as well as X-ray feedback. Here,} $n_\fHI$ is the number density of hydrogen atoms and ions, $h_p$ Planck's constant, $f_{\sm{ion}}$ describes ionization from X-rays and UV photons (see \S \ref{stellar}), and the parametrized case B recombination coefficient for atomic hydrogen $\alpha_\fHI$ is given by
\begin{equation}
\alpha_\fHI=F\times10^{-13}\frac{at^b}{1+ct^d}\ \mathrm{cm}^3\ \mathrm{s}^{-1}
\end{equation} 
with $a=4.309$, $b=-0.6166$, $c=0.6703$, $d=0.5300$ and $t=T/10^4\ K$ {\citep{Pequignot, Hummer} }. This coefficient takes into account that direct recombination into the ground state does not lead to a net increase of neutral hydrogen atoms, since the photon emitted in the recombination process can ionize other hydrogen atoms in the neighbourhood. 

 The fudge factor $F=1.14$ serves to speed up recombination and is determined from comparison with the multilevel-code. The photoionization coefficient $\beta_\fHI$ is calculated from detailed balance at high redshifts as described by \citet{Seager, SeagerFast}. Once the ionized fraction drops below $98\%$, it is instead calculated from the photoionization cross section given by \citet{Sasaki}, as detailed balance may considerably overestimate the photoionization coefficient in case of additional energy input at low redshift.  The wavelength $\lambda_{\fHI, 2p}$ corresponds to the Lyman-$\alpha$ transition from the $2p$ state to the $1s$ state of the hydrogen atom. The frequency for the two-photon transition between the states $2s$ and $1s$ is close to Lyman-$\alpha$ and is thus approximated by $\nu_{\fHI, 2s}=c/\lambda_{\fHI, 2p}$ (i.e., the same averaged wavelength is used). Finally, $\Lambda_\fHI=8.22458\ {\rm s}^{-1}$ is the two-photon rate for the transition $2s$-$1s$ according to \citet{Goldman} and $K_\fHI\equiv \lambda_{\fHI, 2p}^3/[8\pi H(z)]$ the cosmological redshifting of Lyman $\alpha$ photons. To take into account UV feedback, which can not be treated as a background but creates locally ionized bubbles, we further calculate the volume-filling factor $Q_{HII}$ of ionized hydrogen according to the prescriptions of \citep{Shapiro, Haiman, BarkanaR, LoebR, Choudhury, Schneider}. We assume that the temperature in the ionized medium is the maximum between $10^4$ K and the temperature in the overall neutral gas. The latter is only relevant if primordial magnetic fields have heated the medium to temperatures above $10^4$ K before it was ionized.

\subsection{Heating due to primordial magnetic fields}\label{magnetic}
The presence of magnetic fields leads to two different contributions to the heating rate, one coming from ambipolar diffusion and one resulting from the decay of MHD turbulence. In the first case, the contribution can be calculated as \citep{Sethi05, Cowling, Schleicher}:
\begin{equation}
L_{\sm{ambi}}\sim\frac{\rho_n}{16\pi^2\gamma \rho_b^2 \rho_i}\frac{B^4}{L^2}. \label{ambiheatapprox}
\end{equation}
Here, $\rho_n$, $\rho_i$ and $\rho_b$ are the mass densities of neutral hydrogen, ionized hydrogen and all baryons. The ion-neutral coupling coefficient is calculated using the updated zero drift velocity momentum transfer coefficients of \citet{Pinto} for collisions of \HII  with \HI and {\HeI}$\!\!$.
The coherence length $L$ is estimated as the inverse of the Alfv\'{e}n damping wavelength $k_{\sm{max}}^{-1}$  given by \citep{Jedamzik, Subramanian, Seshadri} as
\begin{eqnarray}
k_{\sm{max}}&\sim&234\ \mathrm{Mpc}^{-1}\left(\frac{B_0}{10^{-9}\ \mathrm{G}} \right)^{-1}\left(\frac{\Omega_m}{0.3}\right)^{1/4}\nonumber\\
&\times&\left(\frac{\Omega_b h^2}{0.02}\right)^{1/2}\left(\frac{h}{0.7} \right)^{1/4},\label{minlength}
\end{eqnarray}
which is the scale on which fluctuations in the magnetic field are damped out during recombination. The comoving magnetic field is denoted as $B_0=B/(1+z)^2$ .
\\ \\
For decaying MHD turbulence, we adopt the prescription of \citet{Sethi05}, 
\begin{equation}
L_{\sm{decay}}=\frac{B_0(t)^2}{8\pi}\frac{3\tilde{\alpha}}{2}\frac{[\ln(1+t_d/t_i)]^{\tilde{\alpha}} H(t)}{[\ln(1+t_d/t_i)+\ln(t/t_i) ]^{\tilde{\alpha}+1}},\label{decayheat}
\end{equation}
where $t$ is the cosmological time at redshift $z$, $t_d$ is the dynamical timescale, $t_i$ the time where decay starts, i.\ e.\ after the recombination epoch when velocity perturbations are no longer damped by the large radiative viscosity, $z_i$ is the corresponding redshift. For a power spectrum of the magnetic field with power-law index $\alpha$, {implying that the magnetic field scales with the wavevector $k$ to the power $3+\alpha$,} the parameter $\tilde{\alpha}$ is given as $\tilde{\alpha}=2(\alpha+3)/(\alpha+5)$ \citep{Olesen, Shiromizu, Christensson, Banerjee03}. In the generic case, we expect the power spectrum of the magnetic field to have a maximum at the scale of the coherence length, and the heat input by MHD decay should be determined from the positive slope corresponding to larger scales \citep{Mueller, Christensson, Banerjee03, Banerjee04}. We thus adopt $\alpha=3$ for the calculation. We estimate the dynamical timescale as $t_d=L/v_A$, where $v_A=B/\sqrt{4\pi \rho_b}$ is the Alv\'{e}n velocity and $\rho_b$ the baryon mass density. The evolution of the magnetic field as a function of redshift can be determined from the magnetic field energy density $E_B=B^2/8\pi$, which evolves as \citep{Sethi05}
\begin{equation}
\frac{dE_B}{dt}=-4H(t)E_B-L_{\sm{ambi}}-L_{\sm{decay}}.\label{bfield}
\end{equation}
For the models given in Table \ref{tab:models}, the evolution of temperature and ionization in the overall neutral medium, i. e. the gas that was not yet affected by UV photons from reionization, is given in Fig. \ref{tempion}. 

{For comoving fields weaker than $0.02$~nG, the gas temperature is closely coupled to the CMB via Compton-scattering at redshifts $z>200$. At lower redshifts, this coupling becomes inefficient and the gas cools adiabatically during expansion, until heating due to X-ray feedback becomes important near redshift $20$. The ionized fraction drops rapidly from fully ionized at $z\sim1100$ to $\sim2\times10^{-4}$, until it increases again at low redshift due to X-ray feedback. For larger magnetic fields, the additional heat input becomes significant and allows the gas to decouple earlier from the CMB. For comoving fields of order $0.5$~nG or more, the gas reaches a temperature plateau near $10^4$~K. At this temperature scale, collisional ionizations become important and the ionized fraction in the gas increases, such that ambipolar diffusion heating becomes inefficient. The onset of X-ray feedback is delayed to the higher filtering mass in the presence of magnetic fields, which is discussed in the next section.}

\subsection{The filtering mass scale and stellar feedback}\label{stellar}
The universe becomes reionized due to stellar feedback. We assume here that the star formation rate (SFR) is proportional to the change in the collapsed fraction $f_{\sm{coll}}$, i.\ e.\ the fraction of mass in halos more massive than $m_{\sm{min}}$. This fraction is given by the formalism of \citet{Press}. In \citet{Schleicher}, we have introduced the generalized filtering mass $m_{F,B}$, given as
\begin{equation}
M_{F,B}^{2/3}=\frac{3}{a}\int_0^a da' M_g^{2/3}(a')\left[1-\left(\frac{a'}{a} \right)^{1/2} \right],
\end{equation}
where $a=1/(1+z)$ is the scale factor and $M_g$ is the maximum of the thermal Jeans mass $M_J$ and the magnetic Jeans mass $M_J^B$, the mass scale below which magnetic pressure gradients can counteract gravitational colapse \citep{Sethi05, Subramanian}. The concept of the filtering mass, i.\ e.\ the halo mass for which the baryonic and dark matter evolution can decouple, goes back to \citet{GnedinHui}, see also \citet{Gnedin}. To take into account the back reaction of the photo-heated gas on structure formation, the Jeans mass is calculated from the effective temperature $T_{eff}=Q_{HII}T_{max}+(1-Q_{HII})T$, where $T_{max}=\mathrm{max}(10^4\ \mathrm{K},T)$ \citep{Schneider}. The evolution of the filtering mass for different magnetic field strengths is given in Fig. \ref{filt}. 

{When the comoving field is weaker than $0.03$~nG, the filtering mass is of order $10^5\,M_\odot$ at high redshift and increases up to $10^7\,M_\odot$ during reionization, as UV and X-ray feedback heats the gas temperature and prevents the collapse of baryonic gas on {\em small scales}. Stronger magnetic fields provide additional heating via ambipolar diffusion and decaying MHD turbulence, such that the filtering mass is higher from the beginning. This implies that the first luminous objects form later, as the gas in halos below the filtering mass cannot collapse. This effect is even stronger when the comoving field is stronger than $0.3$~nG, as the magnetic Jeans mass then dominates over the thermal Jeans mass and increases the filtering mass scale by further orders of magnitude. As discussed in \citet{Schleicher}, the feedback on structure formation via the filtering mass can delay reionization significantly, which in turn allows to calculate upper limits on the magnetic field strength due to the measured reionization optical depth. }

{Consequently, we define the lower limit to halo masses that can form stars as} $m_{\sm{min}}={\mathrm{max}}(M_{F,B},10^5\ M_\odot)$, where $10^5\ M_\odot$ is the minimal mass scale for which baryons can cool efficiently, as found in simulations of \citet{Greif}. We take into account X-ray feedback that can penetrate into the IGM due to its long mean-free path, as well as feedback from UV photons produced in the star forming regions. As we showed in \citet{Schleicher}, Population II stars can not reionize the universe efficiently. It is thus reasonable to assume that the first luminous sources were indeed massive Pop. III stars. While this may no longer be true once chemical and radiative feedback becomes efficient and leads to a different mode of star formation, such a transition seems not important for the main purpose of this work, which is to demonstrate how primordial magnetic fields shift the onset of reionization and the epoch where a Lyman $\alpha$ background builds up initially. We thus adopt model B of \citet{Schleicher}, which uses an escape fraction of $100\%$ if the virial temperature is below $10^4$ K, and $10\%$ {for the atomic cooling halos with higher virial temperature. It further assumes that $4\times10^4$ UV photons are emitted per stellar baryon during the lifetime of the star. This choice is in particular motivated by simulations of \citet{Whalen}, that show that minihalos are easily photo-evaporated by UV feedback from Pop. III stars, yielding escape fractions close to $100\%$. A new study by \citet{Wise08} shows that the escape fraction may be larger than $25\%$ even for atomic cooling halos. However, their study assumes purely primordial gas, which is unlikely in such systems, and we expect that studies which take into account metal enrichment will find lower escape fractions, perhaps of the order $10\%$ as suggested here.}

\section{The $21$ cm background}\label{21cm}
The $21$ cm brightness temperature depends on the hyperfinestructure level populations of neutral hydrogen, which is described by the spin temperature $T_{spin}$. In this section, we review the physical processes that determine the spin temperature, discuss the build-up of a Lyman-$\alpha$ background that can couple the spin temperature to the gas temperature at low redshift, and discuss various sources of $21$ cm brightness fluctuations.
\subsection{The spin temperature}
% \begin{figure}
%   \centering
%    \subfigure[]{\includegraphics[scale=0.4]{f3a.eps}\label{example1}}\qquad
%    \subfigure[]{\includegraphics[scale=0.4]{f3b.eps}\label{example2}}\qquad
%   \caption{CMB, gas and spin temperature in the gas unaffected by reionization, for a star formation efficiency $f_*=10^{-3}$, in the zero field case (a) and for a field strength of $0.8$ nG (b). In the absence of magnetic fields, gas and CMB are strongly coupled due to efficient Compton scattering of CMB photons. At about redshift $200$, the gas temperature decouples and evolves adiabatically, until it is reheated due to X-ray feedback. The spin temperature follows the gas temperature until redshift $\sim100$. Collisional coupling then becomes less effective, but is replaced by the Wouthuysen-Field coupling at $z\sim25$. In the presence of magnetic fields, additional heat goes into the gas, it thus decouples earlier and its temperature rises above the CMB. Wouthuysen-Field coupling becomes effective only at $z\sim15$, as the high magnetic Jeans mass delays the formation of luminous sources significantly.
% } 
% \end{figure}

\bef
\showtwover{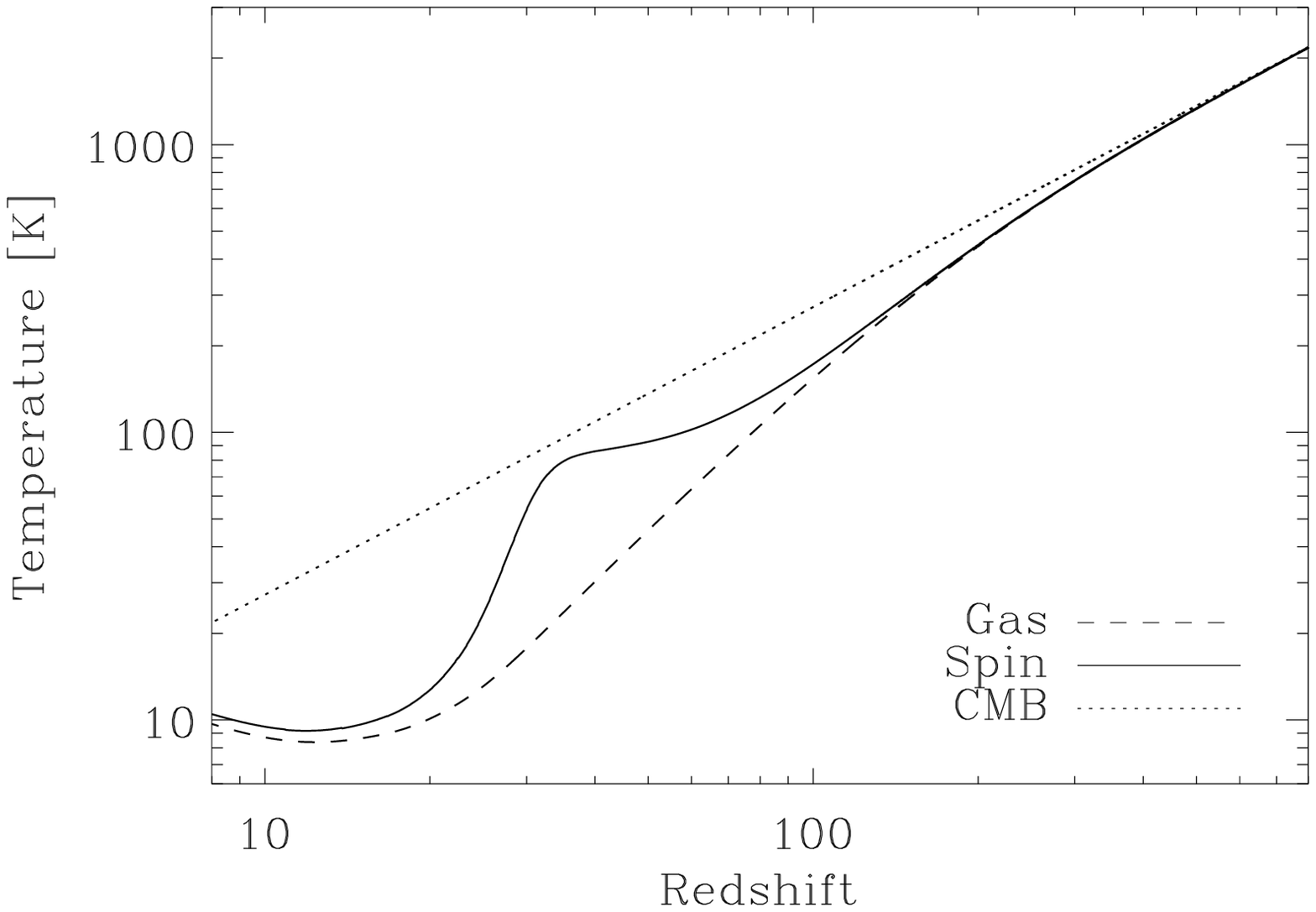}
	{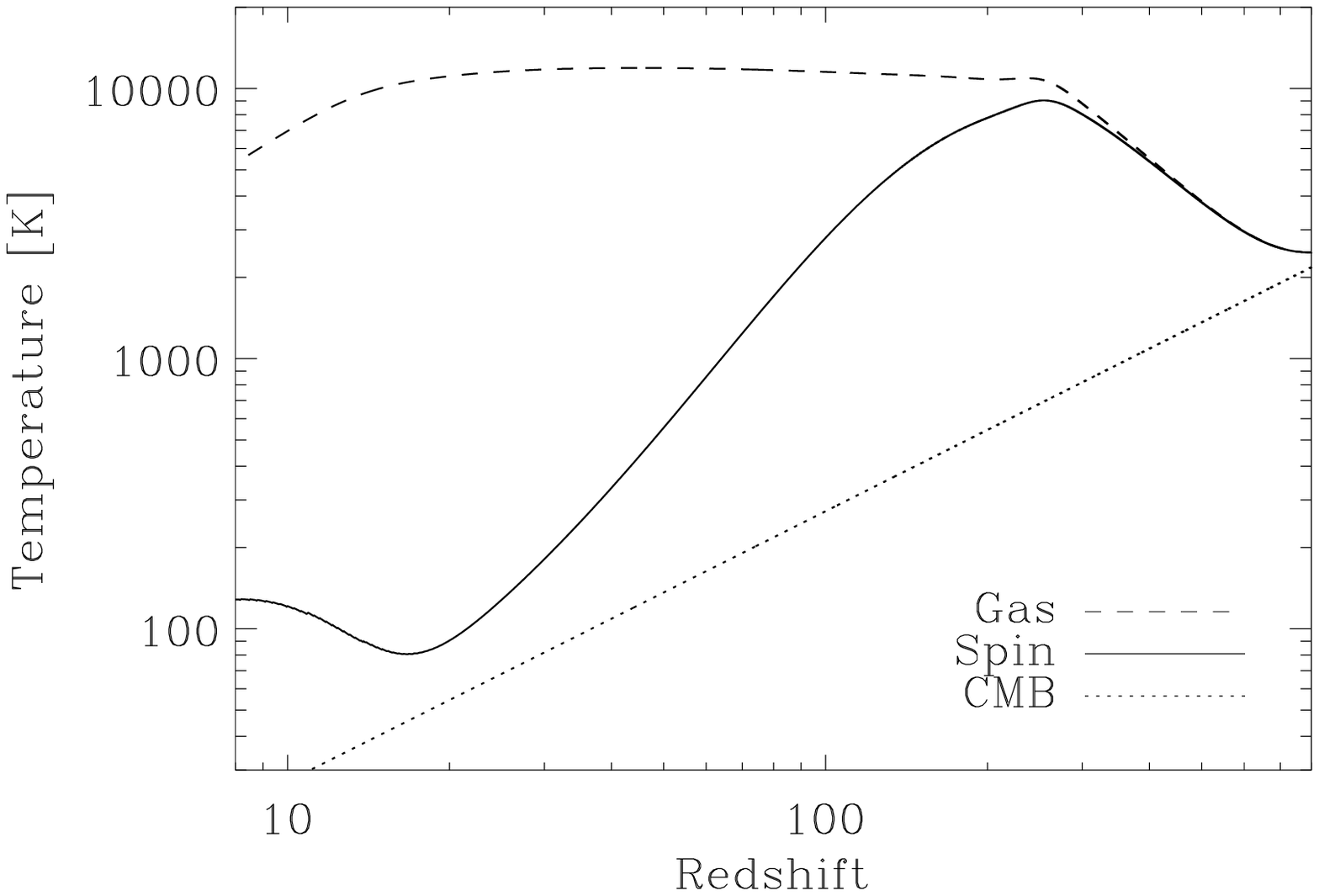}
\caption{CMB, gas and spin temperature in the gas unaffected by reionization, for a star formation efficiency $f_*=10^{-3}$, in the zero field case (upper panel) and for a field strength of $0.8$ nG (lower panel). In the absence of magnetic fields, gas and CMB are strongly coupled due to efficient Compton scattering of CMB photons. At about redshift $200$, the gas temperature decouples and evolves adiabatically, until it is reheated due to X-ray feedback. The spin temperature follows the gas temperature until redshift $\sim100$. Collisional coupling then becomes less effective, but is replaced by the Wouthuysen-Field coupling at $z\sim25$. In the presence of magnetic fields, additional heat goes into the gas, it thus decouples earlier and its temperature rises above the CMB. Wouthuysen-Field coupling becomes effective only at $z\sim15$, as the high magnetic Jeans mass delays the formation of luminous sources significantly.}
\label{example}
\eef

% \begin{figure}
%   \centering
%    \subfigure[]{\includegraphics[scale=0.4]{f6.eps}\label{tspin}}\qquad
%    \subfigure[]{\includegraphics[scale=0.4]{f7.eps}\label{dt21}}\qquad
% \subfigure[]{\includegraphics[scale=0.4]{f8.eps}\label{dtdf}}\qquad
%   \caption{The mean evolution in the overall neutral gas, calculated for the models described in Table \ref{tab:models}. (a) shows the evolution of the spin temperature, (b) the mean brightness temperature fluctuation and (c) the frequency gradient of the mean brightness temperature fluctuation.} 
% \end{figure}

\bef
\showthreeover{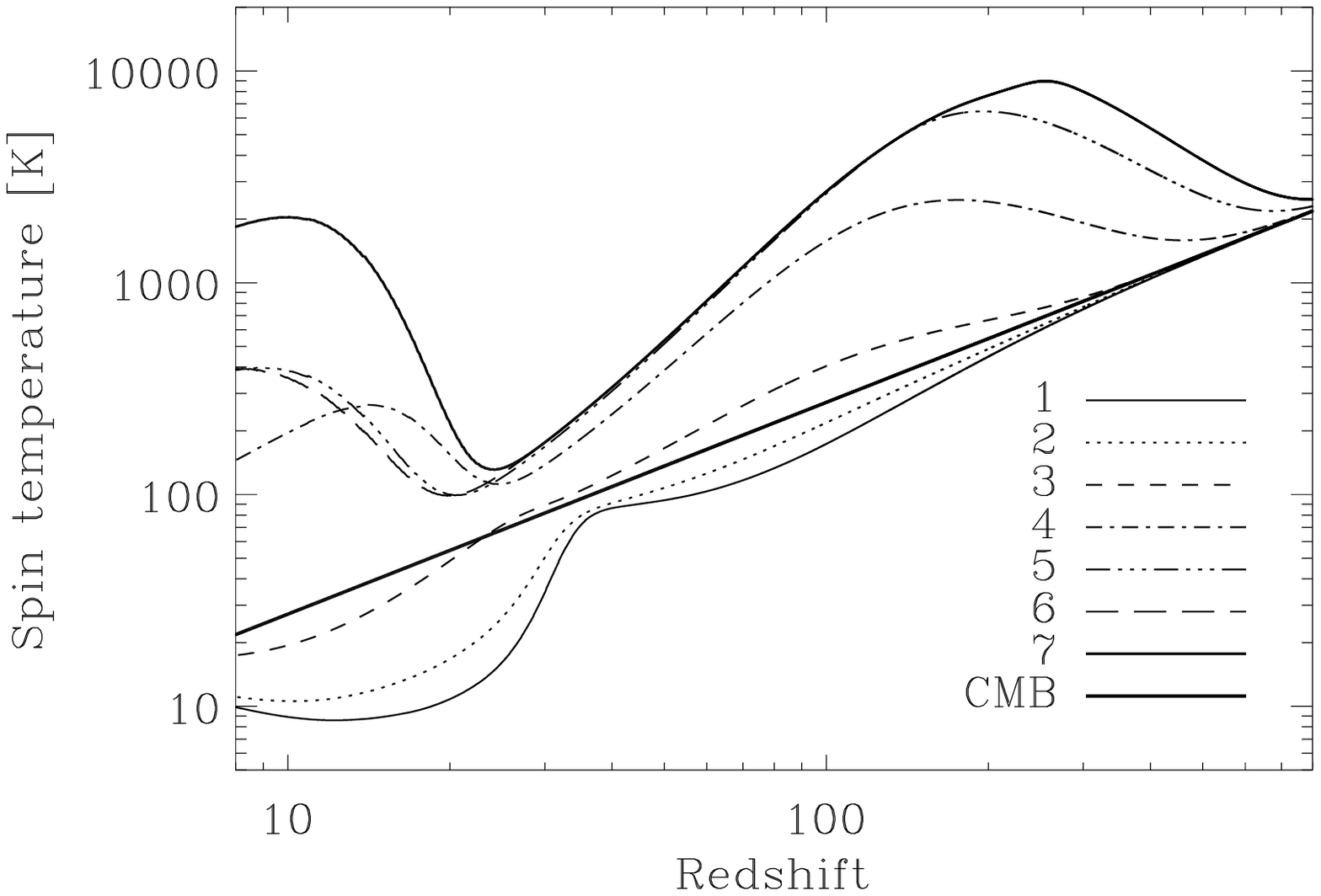}
	{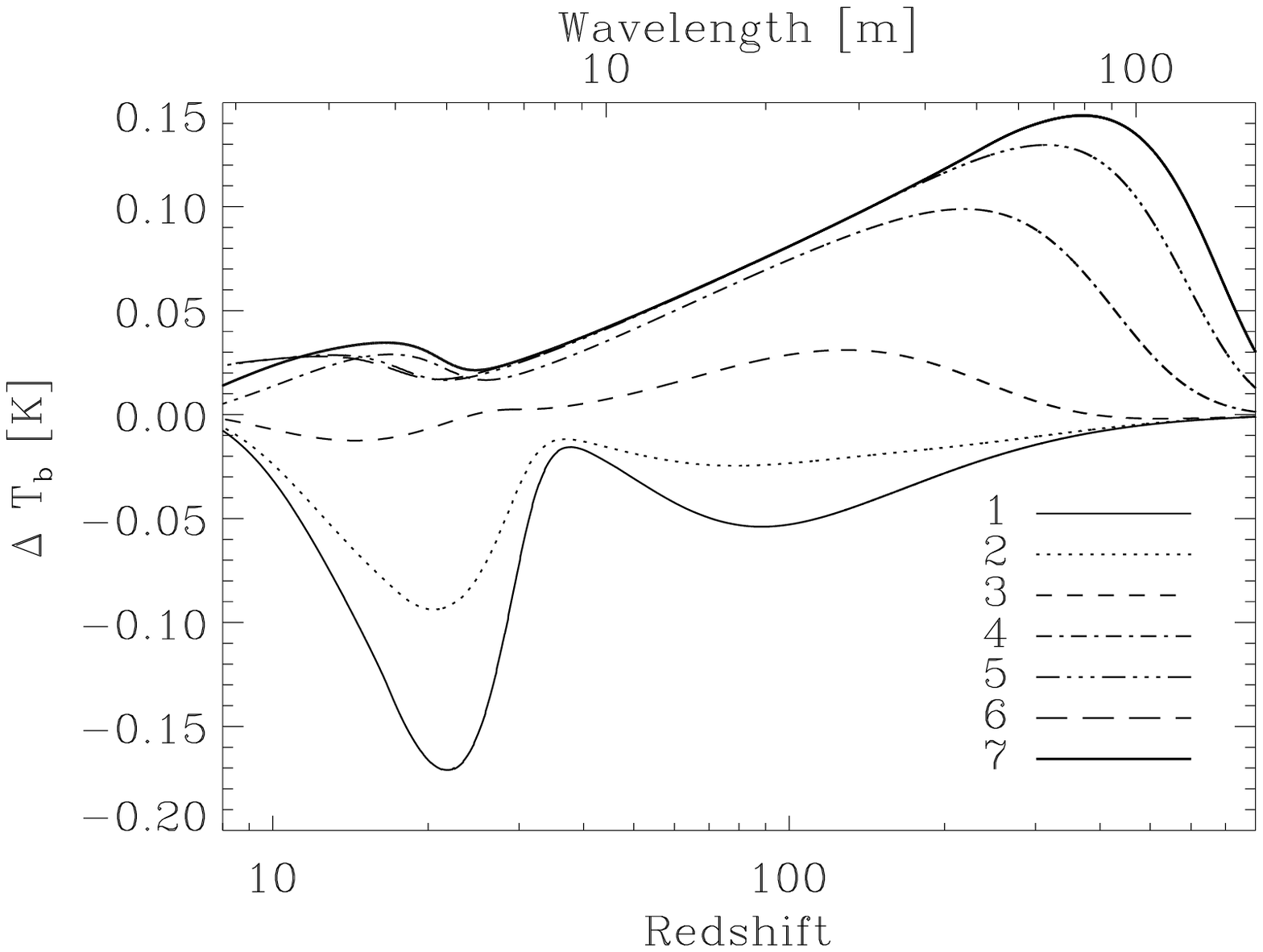}{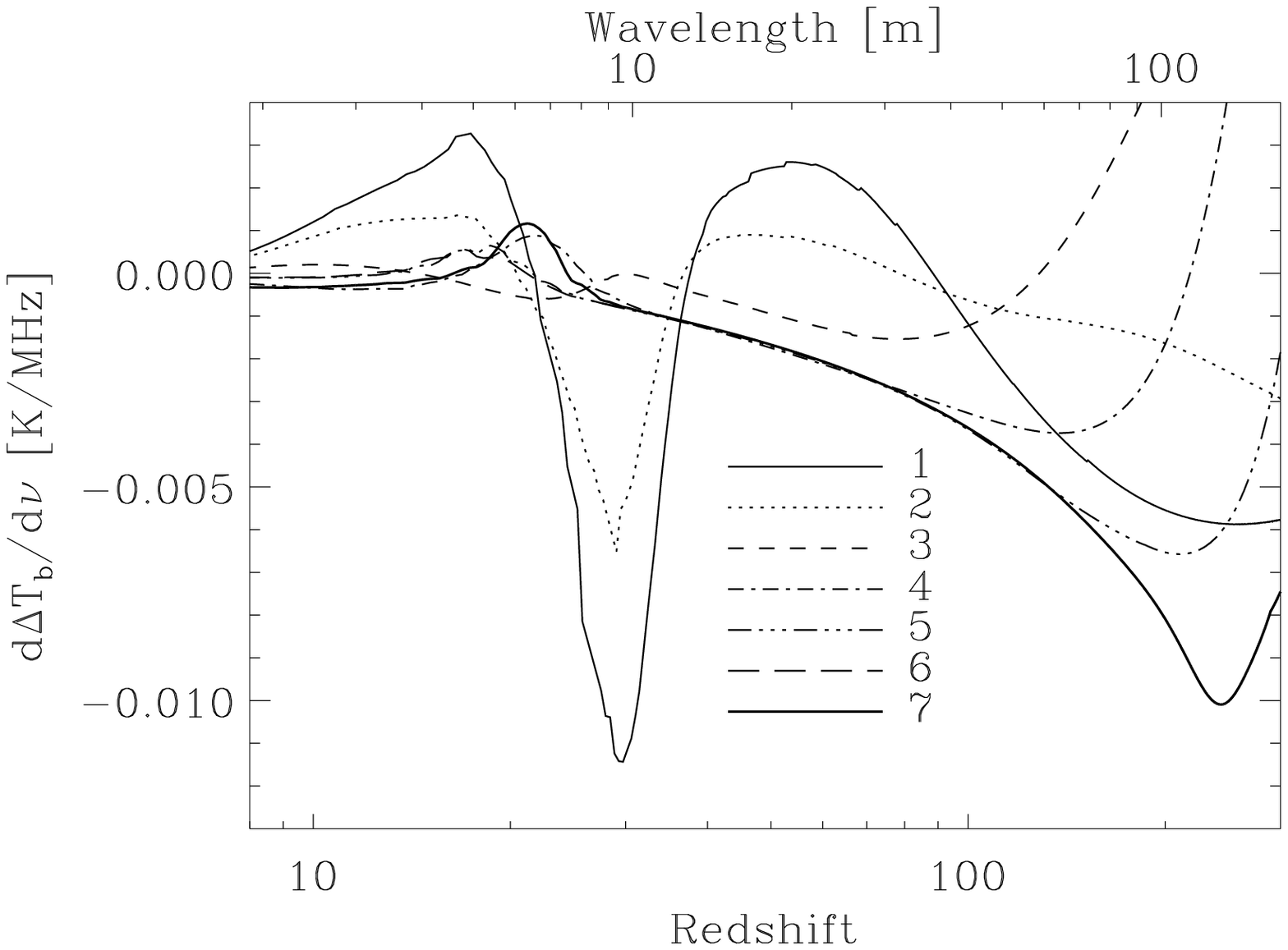}
\caption{The mean evolution in the overall neutral gas, calculated for the models described in Table \ref{tab:models}. Top: The spin temperature. Middle: The mean brightness temperature fluctuation. Bottom: The frequency gradient of the mean brightness temperature fluctuation.}
\label{spin}
\eef

The observed $21$ cm brightness temperature fluctuation can then be conveniently expressed as \citep{Furlanetto}
\begin{eqnarray}
\delta T_b &=& 27 x_{\fHI}(1+\delta)\left(\frac{\Omega_b h^2}{0.023} \right) \left(\frac{0.15}{\Omega_m h^2}\frac{1+z}{10} \right)^{1/2}\nonumber\\
&\times& \left(\frac{T_S-T_r}{T_S}\right) \left(\frac{H(z)/(1+z)}{dv_{||}/dr_{||}} \right) \ \mathrm{mK},
\end{eqnarray}
where $\delta$ is the fractional overdensity, $x_{\fHI}=1-x_{\sm{eff}}$ the effective neutral fraction, $x_{\sm{eff}}=Q_{\fHII}+(1-Q_{\fHII})x_{\fpp}$ the effective ionized fraction, $\Omega_b$ and $\Omega_m$ the cosmological density parameters for baryonic and total mass, and $dv_{||}/dr_{||}$ the gradient of the proper velocity along the line of sight, including the Hubble expansion. The spin temperature $T_{spin}$ is given as.
\begin{equation}
T_{spin}^{-1}=\frac{T_r^{-1}+x_c T^{-1}+x_\alpha T_c^{-1}}{1+x_c+x_\alpha},
\end{equation}
where $T_c$ is the colour temperature, $x_c=x_c^{HH}+x_c^{eH}$ the collisional coupling coefficient, given as the sum of the coupling coefficient with respect to hydrogen atoms, $x_c^{HH}=\kappa_{1-0}^H n_H T_*/A_{10}T_r$, and the coupling coefficient with respect to electrons, $x_c^{eH}=\kappa_{1-0}^e n_e T_*/A_{10}T_r$. Further, $n_H$ denotes the number density of hydrogen atoms, $T_*=0.068\ \mathrm{K}$ the temperature corresponding to the $21$ cm transition and $A_{10}=2.85\times10^{-15}\ \mathrm{s}^{-1}$ the Einstein coefficient for radiative de-excitation. For the collisional de-excitation rate through hydrogen, $\kappa_{1-0}^H$, we use the recent calculation of \citet{Zygelman}, and we use the new rate of \citet{FF07} for the collisional de-excitation rate through electrons, $\kappa_{1-0}^H$. When both the spin and gas temperature are significantly larger than the spin exchange temperature $T_{se}=0.4$ K, as it is usually the case, the colour temperature is given as
\begin{equation}
T_c=T\left(\frac{1+T_{se}/T}{1+T_{se}/T_{spin}} \right).
\end{equation}
Spin and colour temperature thus depend on each other and need to be solved for simultaneously. The Wouthuysen-Field coupling coefficient is given as
\begin{equation}
x_\alpha=S_\alpha\frac{J_\alpha}{J_c},
\end{equation}
where $J_c=1.165\times10^{-10}(1+z)/20\ \mathrm{cm}^{-2}\ \mathrm{s}^{-1}\ \mathrm{Hz}^{-1}\ \mathrm{sr}^{-1}$ and $J_\alpha$ is the proper Lyman $\alpha$ photon intensity, given as 
\begin{equation}
J_\alpha(z)=\sum_{n=2}^{n_{\sm{max}}}f_{\sm{rec}}(n)\int_z^{z_{\sm{max}}}dz'\frac{(1+z)^2}{4\pi}\frac{c}{H(z')}\epsilon(\nu_n',z')
\end{equation}
\citep[see also][]{Barkana3}. $f_{\sm{rec}}$ is the recycling probability of Lyman $n$ photons \citep{Pritchard, Hirata}, $H(z)$ the Hubble function at redshift $z$, $c$ the speed of light, $n_{\sm{max}}$ the highest considered Lyman $n$ resonance, $z_{\sm{max}}$ the highest redshift at which star formation might occur and $\epsilon(\nu,z)$ the comoving emissivity. We adopt $n_{\sm{max}}=30$ and $z_{\sm{max}}=40$. The suppresion factor $S_\alpha$ can be calculated as \citep{Chuzhoy, FP06}
\begin{equation}
S_\alpha=\mathrm{exp}\left(-0.803T^{-2/3}\left(10^{-6}\tau_{\sm{GP}} \right)^{1/3}\right)
\end{equation}
from the optical depth for Lyman $\alpha$ photons,
\begin{equation}
 \tau_{\sm{GP}}\sim3\times10^5 x_{HI}\left(\frac{1+z}{7} \right)^{3/2}.
\end{equation}
Examples for the evolution of the gas, spin and CMB temperature in the zero-field case and for a magnetic field of $0.8$ nG are given in Fig. \ref{example}.

 \subsection{The build-up of a Lyman-$\alpha$ background}\label{Lymanbg}
In the dark ages, the universe is transparent to photons between the Lyman $\alpha$ line and the Lyman limit. Luminous sources can thus create a radiation background between these frequencies that is redshifted during the expansion of the universe. When a photon is shifted into a Lyman line, it is scattered and can couple the spin temperature to the gas temperature as described above. As shown by \citet{Hirata} and \citet{Barkana3}, a population of Pop. III stars can already produce a significant amount of Lyman $\alpha$ radiation around redshift $25$. We assume that emission is proportional to the star formation rate, which is assumed to be proportional to the change in the collapsed fraction $f_{\sm{coll}}$. It thus depends on the evolution of the generalized filtering mass in the presence of magnetic fields. As discussed in \citet{Furlanetto} and \citet{Barkana3}, the comoving emissivity is given as
\begin{equation}
\epsilon(\nu,z)=f_* n_b \epsilon_b(\nu)\frac{df_{\sm{coll}}}{dt},
\end{equation}
where $n_b$ is the comoving baryonic number density and $\epsilon_b(\nu)$ the number of photons produces in the frequency interval $\nu\pm d\nu/2$ per baryon incorporated in stars. The emissivity is approximated as a power law $\epsilon_b(\nu)\propto\nu^{\alpha_s-1}$, and the parameters must be chosen according to the model for the stellar population. As discussed above, we assume that the first stars are massive Pop. III stars, and adopt the model of \citet{Barkana3} with $N_\alpha=2670$ photons per stellar baryon between the Lyman $\alpha$ and the Lyman $\beta$ line and a spectral index $\alpha_s=1.29$
%For models 1 to 4, which assume a spectrum according to the locally measured IMF of low-metallicity stars \citep{Scalo,Kroupa,Chabrier}, we fix them according to \citet{Leitherer} by assuming that Pop. II stars emit $N_\alpha=6520$ photons per baryon between the Lyman $\alpha$ and the Lyman $\beta$ line with a spectral index $\alpha_s=0.14$. For models 5 to 7, which correspond to massive Pop III stars, we assume that the stars emit $N_\alpha=2670$ photons per baryon between the Lyman $\alpha$ and the Lyman $\beta$ line with a spectral index $\alpha_s=1.29$ \citep{Bromm2}. These models have also been adopted by \citet{Barkana3} for the calculation of the Lyman $\alpha$ background. 

In addition, secondary effects from X-ray photons may lead to excitation of hydrogen atoms and the production of further Lyman $\alpha$ photons. We calculate the production of Lyman $\alpha$ photons from the X-ray background for consistency. The contribution from the X-rays is then given as \citep{Chen, Chuzhoy2, Furlanetto}
\begin{eqnarray}
 \Delta x_\alpha^{X-\sm{ray}} & = & 0.05S_\alpha f_X
 \left(\frac{f_{X,\sm{coll}}}{1/3}\frac{f_*}{0.1}\frac{df_{\sm{coll}}/dz}{0.01}
 \right) \nonumber \\ 
  & & \times \left(\frac{1+z}{10} \right)^3.
\end{eqnarray}
We adopt $f_X=0.5$, and calculate $f_{X,\sm{coll}}$ from the fit-formulae of \citet{ShullSteen}. {(Notation: Please recall that $f_{\sm{coll}}$ denotes the fraction of collapsed dark matter, while $f_{X,\sm{coll}}$ denotes the fraction of absorbed X-ray energy which goes into X-rays. The parameter $f_X$ provides an overall normalization for the correlation between the star formation rate and the X-ray luminosity \citep{Koyama}.) } For the models described in Table \ref{tab:models}, we have calculated the evolution of the spin temperature and the mean $21$ cm brightness fluctuation. From an observational point of view, it might be more reasonable to focus on the frequency gradient of the mean $21$ brightness \citep{SchneiderD}. As the frequency dependence of the foreground is known, such an analysis may help to distinguish between the foreground and the actual signal. Results for the spin temperature, the mean $21$ cm brightness fluctuation and its frequency gradient are shown in Fig. \ref{spin}. 

{At redshifts above $200$, the spin temperature is coupled closely to the gas temperature due to collisions with hydrogen atoms. At lower redshifts, the coupling becomes inefficient and the spin temperature evolves towards the CMB temperature, until structure formation sets in and the spin temperature is again coupled to the gas temperature via the Wouthuysen-Field effect. The point where this happens depends on the magnetic field strength, which influences the filtering mass and thus the mass scale of halos in which stars can form. Especially in the presence of strong magnetic fields, this delay in structure formation also makes the production of Lyman $\alpha$ photons less efficient and the departure from the CMB temperature is only weak.}

{This bevahiour is reflected in the evolution of the mean brightness temperature fluctuation. For weak magnetic fields, the gas temperature is colder than the CMB because of adiabatic expansion, so the $21$ cm signal appears in absorption. At redshift $40$ where the spin temperature approaches the CMB temperature, a maximum appears as absorption goes towards zero. Comoving fields of $0.05$~nG add only little heat and essentially bring the gas and spin temperatures closer to the CMB, making the mean brightness temperature fluctuation smaller. For stronger fields, the $21$ cm signal finally appears in emission. Due to very effective coupling and strong temperature differences at redshifts beyond $200$, a pronounced peak appears there in emission. At redshifts between $20$ and $10$ which are more accessible to the next $21$ cm telescopes, magnetic fields generally reduce the expected mean fluctuation, essentially due to the delay in the build-up of a Lyman $\alpha$ background. }

{The evolution of the frequency gradient in the mean brightness temperature fluctuation essentially shows strong minima and maxima where the mean brightness temperature fluctuation changes most significantly. In the case of weak fields, this peak is near redshift 30 when coupling via the Wouthuysen-Field effect becomes efficient. For stronger fields, this peak is shifted towards lower redshifts, thus providing a clear indication regarding the delay of reionization.}

Scattering of Lyman $\alpha$ photons in the IGM does not only couple the spin temperature to the gas temperature, but it is also a potential source of heat. Its effect on the mean temperature evolution has been studied by a number of authors \citep{Madau, Chen, Chen2, Chuzhoy, Ciardi}. While a detailed treatment requires to separately follow the mean temperature of hydrogen and deuterium, a good estimate can be obtained using the approach of \citet{Furlanetto, FP06}, i. e.
\begin{equation}
\frac{2}{3}\frac{L_{\mathrm{heat,Ly}\alpha}}{n_{\fHI}k_B T H(z)} \sim\frac{0.8}{T^{4/3}} \frac{x_{\alpha}}{S_{\alpha}}\left(\frac{10}{1+z} \right).
\end{equation}
For star formation efficiencies of the order $0.1\%$, as they are adopted throughout most of this work, Lyman $\alpha$ heating leads to a negligible temperature change of the order $1$ K.

\begin{table}[htdp]
\begin{center}
\begin{tabular}{ccc}
Model & $B_0$ [nG] & $f_*$  \\
\hline
$1$ & $0$ & $0.1\%$  \\
$2$ & $0.02$ & $0.1\%$ \\
$3$ & $0.05$ & $0.1\%$  \\
$4$ & $0.2$ & $0.1\%$  \\
$5$ & $0.5$ &$0.1\%$ \\
$6$ & $0.8$ & $0.1\%$  \\
$7$ & $0.8$ & $1\%$ \\
\hline
\end{tabular}
\end{center}
\caption{A list of models for different co-moving magnetic fields and star formation efficiencies, which are used in several figures for illustrational purposes. We give the comoving magnetic field $B_0$ and the star formation efficiency $f_*$. {For illustration purposes, all models assume a population of massive Pop. III stars. The amount of Lyman $\alpha$ photons produced per stellar baryon would be larger by roughly a factor of 2 if we were to assume Pop. II stars. Assuming that the same amount of mass goes into stars, the coupling via the Wouthuysen-Field effect would start slightly earlier, but the delay due to magnetic fields is still more significant.} %, which is a rather conservative assumption regarding the production of Lyman $\alpha$ photons.
}
\label{tab:models}
\end{table}%

\subsection{$21$ cm fluctuations}
Fluctuations in the $21$ cm brightness temperature can be caused by the relative density fluctuations $\delta$, relative temperature fluctuations $\delta_T$ and relative fluctuations in the neutral fraction $\delta_H$. Fluctuations in the Lyman $\alpha$ background can be an additional source of $21$ cm fluctuations \citep{Barkana3}, but a detailed treatment of these fluctuations is beyond the scope of this work. These fluctuations are only important in the early stage of the build-up of such a background. Once $x_\alpha$ is significantly larger than unity, the spin temperature is coupled closely to the gas temperature, and small fluctuations in the background will not affect the coupling. We adopt the treatment of \citet{Loeb} to calculate the fractional $21$ cm brightness temperature perturbation, given as
\begin{equation}
\delta_{21}(\vec{k})=(\beta+\mu^2)\delta+\beta_H \delta_H+\beta_T \delta_T,\label{delta}
\end{equation}
where $\mu$ is the cosine of the angle between the wavevector $\vec{k}$ and the line of sight, with the expansion coefficients \citep{Barkana, Furlanetto}
\begin{eqnarray}
\beta &=& 1+\frac{x_c}{x_{\sm{tot}}(1+x_{\sm{tot}})},\\
\beta_H&=&1+\frac{x_c^{HH}-x_c^{eH}}{x_{\sm{tot}}(1+x_{\sm{tot}})},\\
\beta_T&=&\frac{T_r}{T-T_r}+\frac{x_c}{x_{\sm{tot}}(1+x_{\sm{tot}})}\frac{d\ln x_c}{d\ln T}.\label{betat}
\end{eqnarray}
The coefficient $x_{\sm{tot}}$ is given as the sum of the collisional coupling coefficient $x_c$ and the Wouthuysen-Field coupling coefficient $x_\alpha$. 

\section{The evolution of linear perturbations in the dark ages}\label{fluctuations}
To determine the fluctuations in the $21$ cm line, we need to calculate the evolution of linear perturbations during the dark ages. We introduce the relative perturbation of the magnetic field, $\delta_{B}$, and the relative perturbation to the ionized fraction, $\delta_i$, which is related to the relative perturbation in the neutral fraction, $\delta_{\sm{H}}$, by $\delta_i=-\delta_{\sm{H}}(1-x_i)/x_i$. In the general case, different density modes trigger independent temperature and ionization fluctuations. This introduces a non-trivial scale dependence \citep{Barkana2, Naoz}. However, on the large scales which might ultimately be observeable, the growing density mode dominates \citep{Bharadwaj}. {The temperature and ionization fluctuations $\delta_T$ and $\delta_i$ are thus related to the density fluctuations $\delta$ by $\delta_T=g_T(z)\delta$ and $\delta_i=g_i(z)\delta$ with redshift-dependent coupling factors $g_i(z)$ and $g_T(z)$.} The time evolution of $g_T$ and $g_i$ is well-known in the absence of magnetic fields \citep{Naoz, Bharadwaj} and has also been studied for the case of dark matter annihilation and decay \citep{FurlanettoDM}. We further introduce {the relative fluctuation in the magnetic field,} $\delta_{B}=g_{B}\delta$. To extend the previous analysis to the situation considered here, we must consider the effect of the new terms due to ambipolar diffusion heating and decaying MHD turbulence in eq. (\ref{temp}). As it has a complicated dependence on different quantities, we will keep this discussion rather generic, so that it can also be applied to other situations as well. 

Let us consider a source term of the form
\begin{equation}
\left(\frac{\delta T}{\delta z} \right)_{\sm{source}}=f(\rho,T,x_i, B).
\end{equation}
{Here, $f(\rho,T,x_i, B)$ corresponds to the term describing ambipolar diffusion heating and decaying MHD turbulence in Eq. (\ref{temp}).}
When the quantities $\rho$, $T$, $x_i$ and $B$ are perturbed, it is straightforward to show that this introduces further terms for the evolution of the temperature perturbation, which are of the form
\begin{eqnarray}
\frac{\delta \delta_T}{\delta z}&=&T^{-1}\big(\delta\rho\frac{\partial f}{\partial \rho}+T\delta_T\frac{\partial f}{\partial T}+\delta_i x_i\frac{\partial f}{\partial x_i}\nonumber\\
&+& \delta_{B}B\frac{\partial f}{\partial B}-f\delta_T \big).\label{newterms}
\end{eqnarray}
{These new terms describe the change in the relative temperature fluctuation, depending on the additional source term $f(\rho,T,x_i, B)$ and its derivatives with respect to gas density, temperature, ionized fraction and magnetic field. The behaviour is dominated by the properties of the ambipolar diffusion heating term.}
The time evolution of the coupling factors $g_T$ and $g_i$ is thus given as
\begin{eqnarray}
\frac{dg_T}{dz}&=&\frac{g_T-2/3}{1+z}+\frac{x_i}{\eta_1 t_\gamma}\frac{g_T T_r-g_i(T_r-T)}{T(1+z)H(z)}\nonumber\\
&+&T^{-1}\big(\rho\frac{\partial f}{\partial \rho}+Tg_T\frac{\partial f}{\partial T}+g_i x_i\frac{\partial f}{\partial x_i}\nonumber\\
&+&g_{B}B\frac{\partial f}{\partial B}-fg_T \big),\label{gtdz}\\
\frac{dg_i}{dz}&=&\frac{g_i}{1+z}+\frac{\alpha_H x_i n_H(1+g_i+\alpha_H' g_T)}{(1+z)H(z)}\label{gidz}.
\end{eqnarray}
{The first term in Eq. (\ref{gtdz}) describes the evolution towards an adiabatic state in the absence of heating or cooling mechanisms. The second term describes the interaction with the CMB via Compton scattering, which in general drives the gas towards an isothermal state, as the heat input per baryon is constant. The further terms are those derived in Eq. (\ref{newterms}). In Eq. (\ref{gidz}), the first term holds the relative fluctuation $\delta_i$ constant in the absence of recombinations, while the second term describes the effect of hydrogen recombination, which tends to drive the corresponding coupling factor $g_i$ towards $-1$. Fluctuations in density and ionization are then anticorrelated.}

This system must be closed with an additional assumption for the fluctuation in the magnetic field. The dominant mechanism for energy loss is ambipolar diffusion, which scales with $\rho_b^{-2}$. It seems thus unlikely that the magnetic field will be dissipated in regions of enhanced density. We rather assume that the magnetic flux is approximately conserved. In the case of dynamically weak field strength, the corresponding surface scales as $\rho^{-2/3}$, so the magnetic field scales as $\rho^{2/3}$, thus $g_B\sim2/3$. Ambipolar diffusion, the dominant new heating term, scales as $\rho^{-2}$. To relate the change in energy to the change in temperature, it is further multiplied by $3/(2nk_B)$, such that in this case $f\propto\rho^{-3}$. This behavior is not fully compensated by the other terms, thus heating is less effective in overdense regions, and $g_T$ tends to drop below the adiabatic value of $2/3$. For the models given in Table \ref{tab:models}, the evolution of $g_T$ and $g_i$ is given in Fig. \ref{spektrum}. 

{In the zero or weak magnetic field case, the coupling factor $g_T$ evolves from zero (efficient coupling to the CMB) towards $2/3$ (adiabatic), until X-ray feedback sets in. For comoving fields larger than $0.05$~nG, there is an early phase around $z\sim500$ where it evolves towards adiabatic behaviour, as the gas recombines more efficiently in these early times and higher densities, where the factor $g_i$ is still of order $-1$. Thus, ambipolar diffusion in density enhancements is more efficient and effectively heats the gas at these redshifts. Below redshift $500$, the relative fluctuation in ionization evolves towards zero, as the gas becomes thinner and hotter. That leads to a phase in which the relative temperature fluctuations evolves towards an isothermal state ($g_T=0$). At low redshifts, X-ray feedback leads to a decrease of the coupling factor $g_T$. Its behaviour is no longer adiabatic because of this additional heat input. The increase in the mean ionized fraction due to secondary ionizations leads to a more negative $g_i$.}

\section{The $21$ cm power spectrum}\label{power}

% \begin{figure}
%   \centering
% \subfigure[]{\includegraphics[scale=0.4]{f9.eps}\label{spektrum}}\qquad
%    \subfigure[]{\includegraphics[scale=0.4]{f10.eps}\label{flucttemp}}\qquad
%    \subfigure[]{\includegraphics[scale=0.4]{f11.eps}\label{fluction}}\qquad
%   %\subfigure[]{\includegraphics[scale=0.4]{dtdf.eps}\label{dtdf}}\qquad
%   \caption{The evolution of the $21$ cm large-scale power spectrum and the growth of large-scale fluctuations in temperature and ionization for the models given in Table \ref{tab:models}. (a) The evolution of the $21$ cm power spectrum, normalized with respect to the baryonic power spectrum at redshift 0. (b) The evolution of the ratio $g_T=\delta_T/\delta$. (c) The evolution of the ratio $g_i=\delta_{x_i}/\delta$.} 
% \end{figure}

\bef
\showthreeover{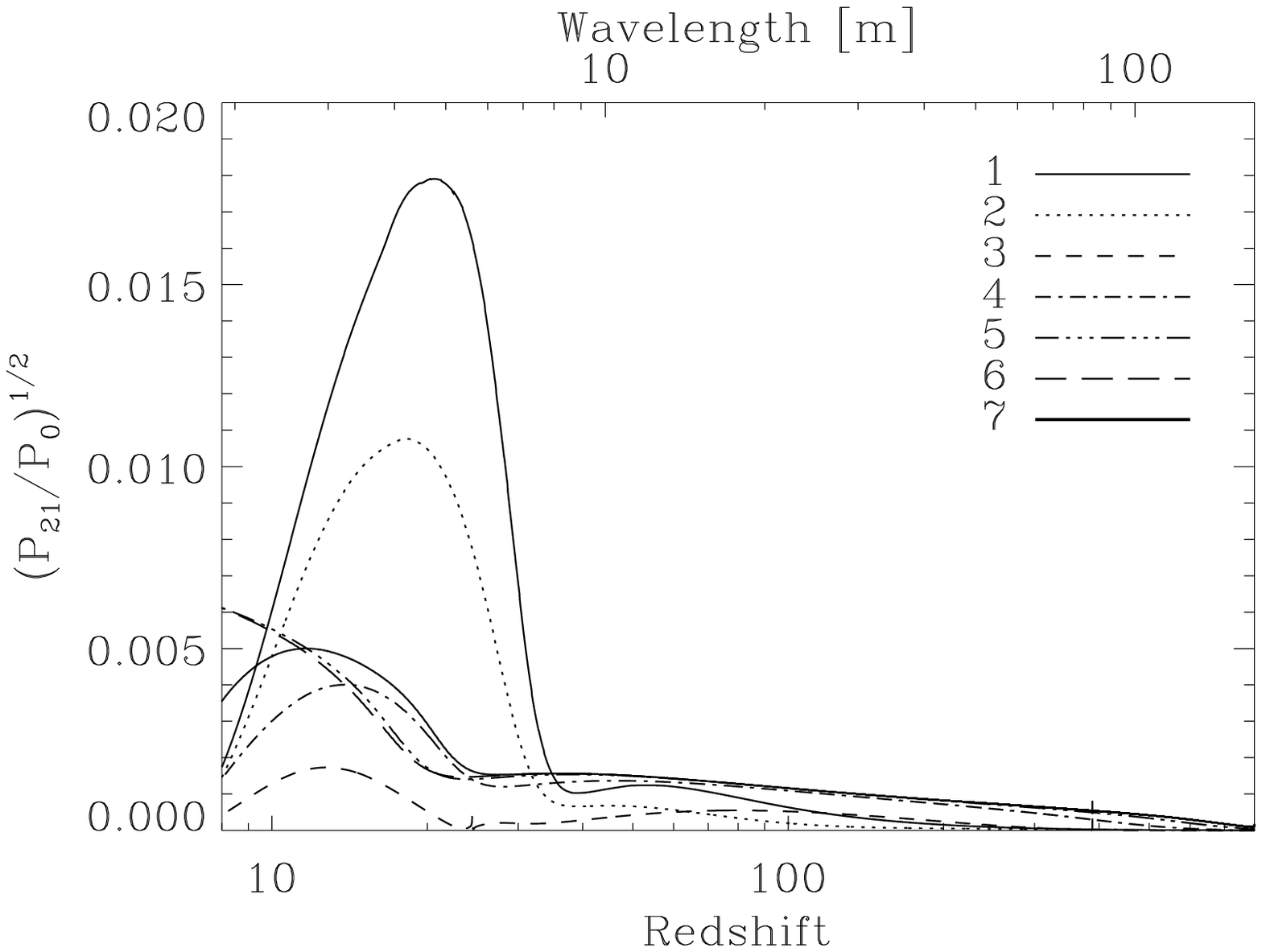}
	{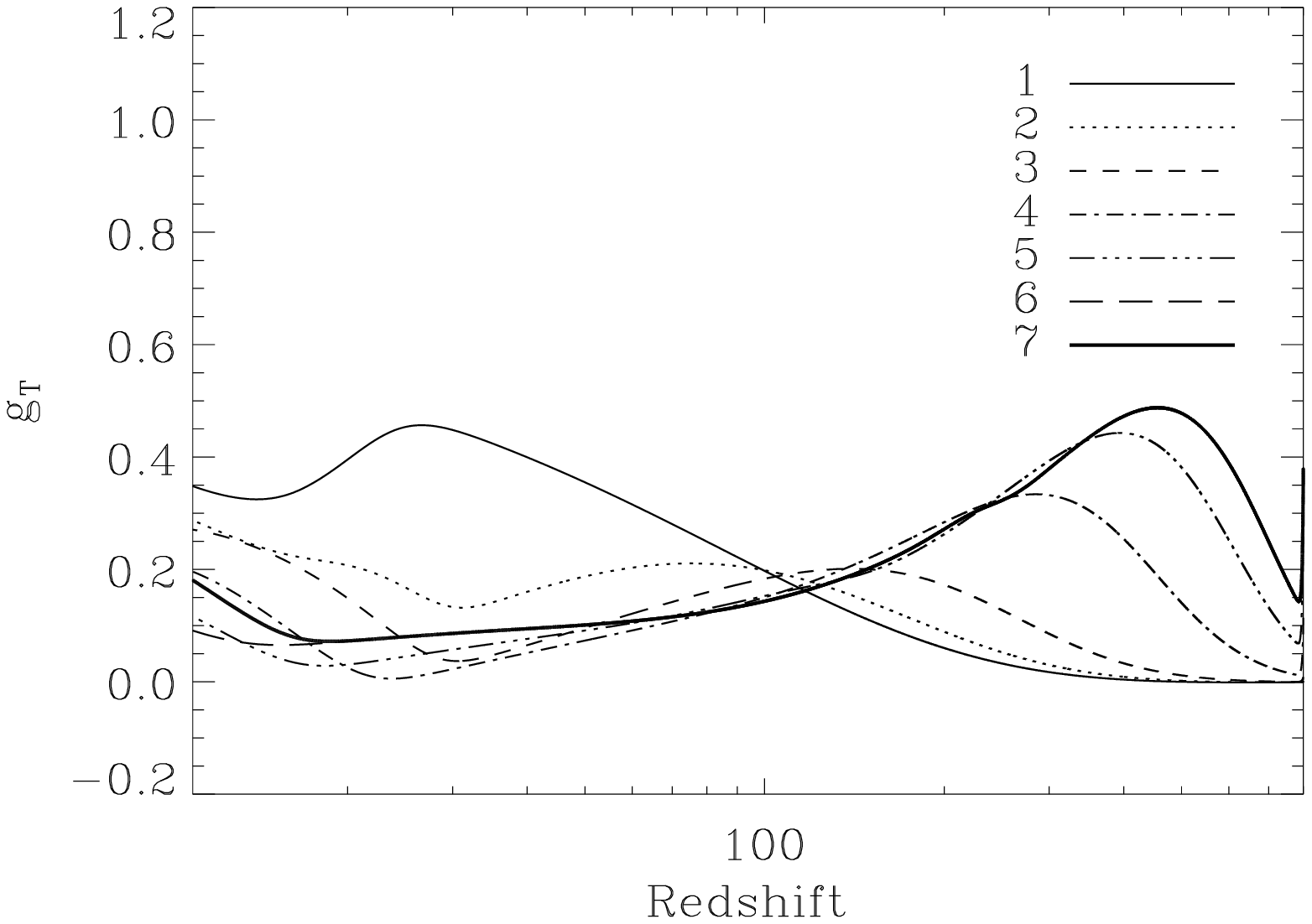}{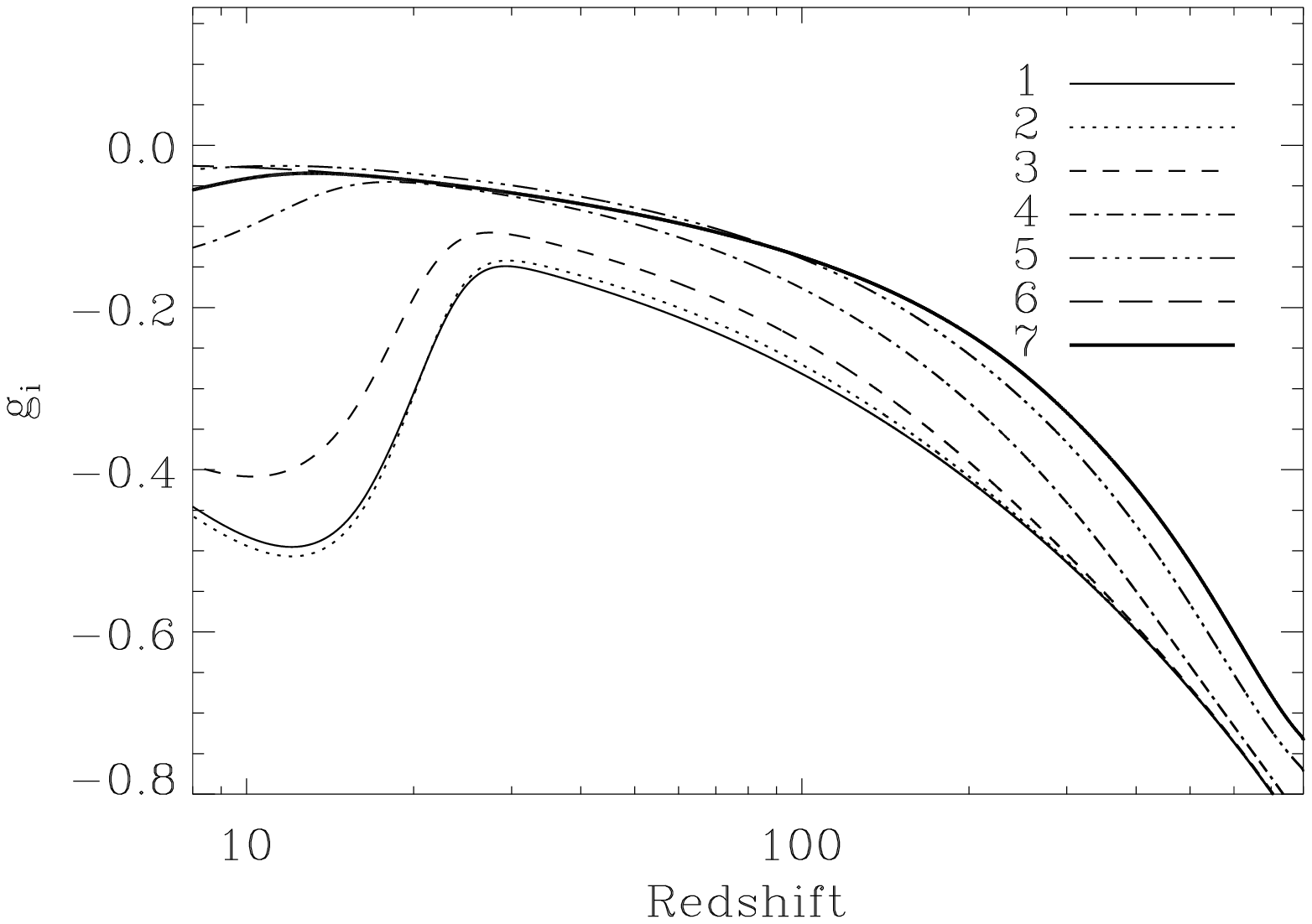}
\caption{The evolution of the $21$ cm large-scale power spectrum and the growth of large-scale fluctuations in temperature and ionization for the models given in Table \ref{tab:models}. Top: The evolution of the $21$ cm power spectrum, normalized with respect to the baryonic power spectrum at redshift 0. Middle: The evolution of the ratio $g_T=\delta_T/\delta$. Bottom: The evolution of the ratio $g_i=\delta_{x_i}/\delta$.}
\label{spektrum}
\eef

%\begin{figure}
%  \centering
%  {\includegraphics[scale=0.45]{power.eps}}
%\caption{The evolution of the power spectrum on large scales as a function of redshift. The model parameters are given in Table \ref{tab:models}. At high redshifts, the signal is increased due to the additional heat in the IGM, which makes collisions more effective and increases the difference between the spin temperature and the CMB temperature. At lower redshifts, the formation of luminous objects and the build-up of a Lyman $\alpha$ background is delayed, tending to decrease the $21$ cm signal compared to the standard case.  }\label{spektrum}
%\end{figure}
We can now calculate the $21$ cm power spectrum in the presence of primordial magnetic fields from eq. (\ref{delta}), leading to 
\begin{equation}
P_{21}(k,\mu)=\delta T_b^2 (\beta'+\mu^2)^2 P_{\delta\delta}(k),
\end{equation}
where $P_{\delta\delta}(k)$ is the baryonic power spectrum and
\begin{equation}
\beta'=\beta+\beta_T g_T-\beta_H x_i g_i/(1-x_i).
\end{equation}
For simplicity, we average over the $\mu$ dependence and define $P_{\delta\delta}=P_{0}/(1+z)^2$, such that
\begin{equation}
P_{21}(k)=\delta T_b^2 \left(\beta'^2+\frac{2}{3}\beta'+\frac{1}{5}\right)\frac{P_{0}(k)}{(1+z)^2}.
\end{equation}
On the large scales considered here, the power spectrum of the $21$ cm line is proportional to the baryonic power spectrum $P_{0}(k)$ for a given redshift, but the proportionality constant will evolve with redshift{ and is independent of the wavenumber in the linear approximation}. The ratio $P_{21}/P_{0}$ is shown in Fig. \ref{spektrum}. {It essentially reflects the behaviour that was found earlier for the mean brightness temperature fluctuation. At high redshift, the amplitude of the $21$~cm power spectrum is larger in the presence of additional heat from ambipolar diffusion and decaying MHD turbulence. At low redshifts, it is smaller in the presence of magnetic fields, as the high filtering mass delays the build-up of a Lyman $\alpha$ background and leads only to a weak coupling of the spin temperature to the gas temperature. In addition, the point where this coupling starts is shifted towards lower redshifts. This provides two signatures that can provide strong indications for the presence of primordial magnetic fields.} 

At high redshifts $z>50$, the additional heat increases the difference between gas and radiation temperature, and the gas decouples earlier from radiation. All that tends to increase the $21$ cm signal at high redshift. For redshifts $z<40$, the situation is more complicated. The altered evolution of the fluctuations and the delay in the formation of luminous objects decrease the amplitude of the power spectrum in total, but also the onset of efficient coupling via the Wouthuysen-Field effect can be shifted significantly. \\ \\
In this context, one might wonder whether $21$ cm observations can actually probe additional small-scale power, which is predicted in the context of primordial magnetic fields \citep{Kim, Sethi05, Tashiro}. However, as discussed for instance by \citet{FurlanettoDM}, the first $21$ cm telescopes will focus on rather larger scales of the order $1$ Mpc. The contribution to the power spectrum from primordial magnetic fields depends on the assumed power spectrum for the magnetic field. Regarding the formation of additional structures, the most significant case is a single-scale power spectrum. In this case, the contribution to the matter power spectrum can be evaluated analytically, yielding \citep{Kim, Sethi05}
\begin{equation}
P(k)=\frac{B_0^4 k^3 H_0^{-4}}{(8\pi)^3\Omega_m^4\rho_c^2k_{max}^2}
\end{equation}
for $k\leq2k_{max}$, where $k_{max}$ is given in Eq. \ref{minlength}. For interesting field strengths of the order of $1$ nG, we thus expect modifications on scales of the order $1$ kpc. For weaker fields, $k_{max}^{-1}$ is shifted to larger scales, but the power spectrum decreases with $B_0^4$. Detecting such changes in the matter power spectrum is thus certainly challenging. {However, the calculation of the filtering mass indicates that the baryonic gas will not collapse in minihalos of small masses when primordial fields are present. Star formation and HII regions from UV feedback are thus limited to more massive systems. This might be an additional way to distinguish the case with and without magnetic fields.}

\section{Discussion and outlook}\label{discussion}
In the previous sections, we have presented a semi-analytic model describing the post-reionization universe and reionization in the context of primordial magnetic fields, and calculated the consequences for the mean $21$ cm brightness fluctuation and the large-scale power spectrum.  formation of 
We identify two regimes in which primordial magnetic fields can influence effects measured with $21$ cm telescopes. At low redshifts, primordial magnetic fields tend to delay reionization and the build-up of a Lyman $\alpha$ background, thus shifting the point where the signal is at its maximum, and changing the amplitude of the $21$ cm power spectrum. The first $21$ cm telescopes like LOFAR and others will focus mainly on the redshift of reionization and can thus probe the epoch when a significant Lyman $\alpha$ background builds up. As our understanding of the first stars increases due to advances in theoretical modeling or due to better observational constraints, this may allow us  to determine whether primordial magnetic fields are needed to delay the build-up of Lyman $\alpha$ photons or not. As demonstrated above, comoving field strengths of the order $1$ nG can delay the build-up of a Lyman $\alpha$ background by $\Delta z\sim 10$, which is significantly stronger than other mechanisms that might delay the formation of luminous objects. Lyman Werner feedback is essentially self-regulated and never leads to a significant suppression of star formation \citep{Johnson, JohnsonO}, X-ray feedback from miniquasars is strongly constrained from the observed soft X-ray background \citep{Salvaterra}, and significant heating from dark matter decay or annihilation would be accompanied by a significant amount of secondary ionization, resulting in a too large optical depth. On the other hand, an important issue is the question regarding the first sources of light. As shown in \citet{Schleicher}, massive Pop. III stars are needed to provide the correct reionization optical depth. On the other hand, an additional population of less massive stars with an IMF according to \citep{Scalo,Kroupa,Chabrier} might be present. Such a population would emit more photons between the Lyman $\alpha$ and $\beta$ line per stellar baryon \citep{Leitherer}, and could thus shift the build-up of a Lyman $\alpha$ background to an earlier epoch. The onset of efficient coupling via the Wouthuysen-Field effect thus translates into a combined constraint on the stellar population and the strength of primordial magnetic fields.\\ \\

A further effect occurs at high redshifts, where the additional heat input from magnetic fields due to ambipolar diffusion and the decay of MHD turbulence increases the $21$ cm signal. As gas decouples earlier from the radiation field, the difference between gas and radiation temperature is larger and collisions are more effective in coupling the spin temperature to the gas temperature. The determination of the $21$ cm signal from this epoch is certainly challenging, as the foreground emission corresponds to temperatures which are higher than the expected $21$ cm brightness temperature by several orders of magnitude. However, as pointed out in other works \citep{DiMatteo, OhMack, Zaldarriaga, Sethi, Yu}, the foreground emission is expected to be featureless in frequency, which may allow for a sufficiently accurate subtraction. In this context, it may also to help to focus on the frequency gradient of the mean brightness temperature, rather than the $21$ cm brightness temperature itself. Upcoming long-wavelength experiments such as LOFAR, 21CMA (former PAST)\footnote{http://web.phys.cmu.edu/~past}, MWA\footnote{http://web.haystack.mit.edu/arrays/MWA/index.html}, LWA\footnote{http://lwa.unm.edu} and SKA\footnote{http://www.skatelescope.org}  may thus detect the additional heat from primordial magnetic fields in the neutral gas, or otherwise set new upper limits on primordial magnetic fields, perhaps down to $B_0\sim0.1$ nG. Like the 21 cm transition of hydrogen, rotational and ro-vibrational transitions of primordial molecules may create interesting signatures in the CMB as well\citep{Schleicher2}, which may provide a further test of the thermal evolution during the dark ages.

%In case of a positive detection, one might even try to discriminate between ambipolar diffusion heating and other energy injection models like dark matter decay or annihilation, as these processes would tend to inject the energy at a later time \citep{FurlanettoDM}. However, that would require to measure the $21$ cm signal at very low frequencies corresponding to very high redshifts. This can only be achieved in future satellite missions. 

\acknowledgments

We thank Benedetta Ciardi, Paul Clark, Christoph Federrath, Andrea Ferrara, Daniele Galli, Simon Glover, Thomas Greif, Antonella Maselli and Francesco Palla for many interesting discussions on the topic. DRGS thanks the Heidelberg Graduate School of Fundamental Physics (HGSFP) and the LGFG for financial support. DRGS further thanks the MPA Garching for financial support during a research visit, and especially Benedetta Ciardi and Antonella Maselli for being very kind host researchers. The HGSFP is funded by the Excellence Initiative of the German Government (grant number GSC 129/1). RB is funded by the Emmy Noether grant (DFG) BA 3607/1. RSK thanks for support from the Emmy Noether grant KL 1358/1. All authors also acknowledge subsidies  from the DFG SFB 439 {\em Galaxies in the Early Universe}. {We thank the anonymous referee for valuable comments that improved our manuscript.}

\clearpage

%% the file, uncomment the \plotone, \plottwo, and
%% If you need a layout that cannot be achieved with \plotone or
%% \plottwo, you can invoke the graphicx package directly with the
%% \includegraphics command or use \plotfiddle. For more information,
%% please see the tutorial on "Using Electronic Art with AASTeX" in the
%% documentation \S at the AASTeX Web site,
%% http://www.journals.uchicago.edu/AAS/AASTeX.
%%
%% The examples below also include sample markup for submission of
%% supplemental electronic materials. As always, be sure to check
%% the instructions to authors for the journal you are submitting to
%% for specific submissions guidelines as they vary from

%% This example uses \plotone to include an EPS file scaled to
%% 80% of its natural size with \epsscale. Its caption
%% has been written to indicate that additional figure parts will be
%% available in the electronic journal.

%\begin{figure}
%\epsscale{.80}
%\plotone{f1.eps}
%\caption{blub\label{fig1}}
%\end{figure}

\clearpage

\end{document}